\documentclass[prd,nofootinbib,showpacs]{revtex4}
\usepackage{epsfig,bm}

\begin{document}

\newcommand{\be}{\begin{equation}}
\newcommand{\ee}{\end{equation}}
\newcommand{\tr}{\mbox{Tr}}
\renewcommand{\vec}[1]{\mathbf{#1}}

\title{Detectability of Microwave Background Polarization}
\author{Emory F. Bunn}
 \email{ebunn@richmond.edu}
 \affiliation{Physics Department, University of Richmond,
Richmond, VA  23173}
\date{\today}

\begin{abstract}
\vspace{\baselineskip}
\vspace{\baselineskip}
Measurement of the amplitudes
of both the E and B components of the CMB polarization will open
new windows onto the early Universe.  Using a
Fisher-matrix formalism, we calculate the required
sensitivities and observing times for an experiment to
measure the amplitudes of both E and B components as
a function of sky coverage, taking full account of the
fact that the two components cannot be perfectly separated
in an incomplete sky map.  We also present a simple
approximation scheme that accounts for mixing
of E and B components in computing predicted errors in
the E-component power spectrum amplitude.  
In an experiment with small sky coverage,
mixing of the two components increases the difficulty
of detecting the subdominant B component by a factor
of two or more in observing time; however, for larger
survey sizes the effect of mixing is less pronounced.
As a result, the
optimal experimental setup for detecting
the B component must cover an area of sky significantly
larger than is found when mixing is neglected.
Surprisingly, mixing of E and B components can enhance
the detectability of the E component by increasing the effective
number of independent modes that probe this component.
The formalism presented in this paper can be used to explore
ways in which
survey geometry and nonuniform noise due to uneven
sky coverage will affect detectability
of the two components.
\end{abstract}
\pacs{98.70.Vc,98.80.Es,98.80.Cq,98.80.-k}

\maketitle

\section{Introduction}

The cosmic microwave background radiation (CMB) contains a wealth of
information about the high-redshift Universe.  Efforts to map the
temperature anisotropy and estimate its power spectrum
have already revolutionized our understanding of
cosmology, strongly suggesting
that the Universe is spatially flat and
constraining other cosmological parameters (\textit{e.g.},
\cite{boom,max,dasi,stompor,bond,jaffe,TZ}).
Future anisotropy measurements, especially the MAP \cite{map}
and Planck \cite{planck} satellite missions, will allow
the cosmological parameters to be determined with unprecedented
precision  (\textit{e.g.}, \cite{cmbpar1,cmbpar2}).

Polarization will be the next frontier in the study of the CMB;
considerable effort is already underway (\textit{e.g.},
\cite{Stag,Hedm,Pete}) to map the CMB polarization structure.
That the CMB should be polarized is an extremely robust prediction
\cite{Rees,Koso}
of the gravitational instability paradigm
of standard cosmological theory,
so detection of CMB polarization
will provide a reassuring confirmation of the standard model.  Moreover,
a large amount of additional information
is expected to be lurking in CMB polarization
maps.  Measurement of the CMB polarization power spectra will
help in breaking certain degeneracies between cosmological parameters
\cite{ZalSperSel,EisHuTeg,BucMooTur,Kin}.
But perhaps most exciting of all is the possibility that CMB polarization
measurements will contain the signature of primordial gravity waves
(tensor perturbations)
produced during an inflationary epoch
\cite{ZalSel,KamKosSte,HuSelWhiZal}.  If these hopes are realized,
we will have far more direct evidence than we currently
do that inflation actually occurred, and we will have probed
the Universe at far earlier times than any current observations.

\begin{figure}[t!]
\centerline{\epsfxsize 3in\epsfbox{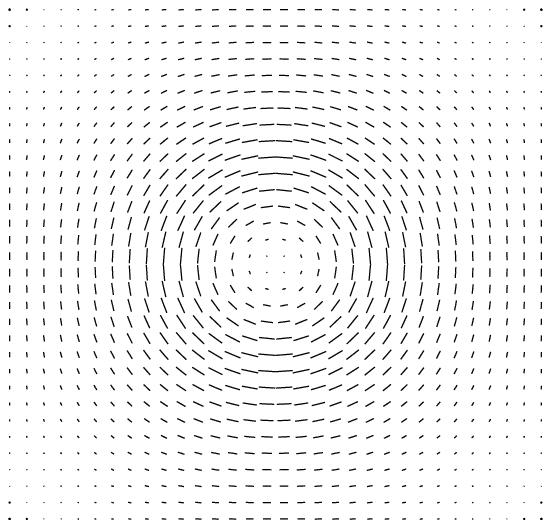}
\epsfxsize 3in\epsfbox{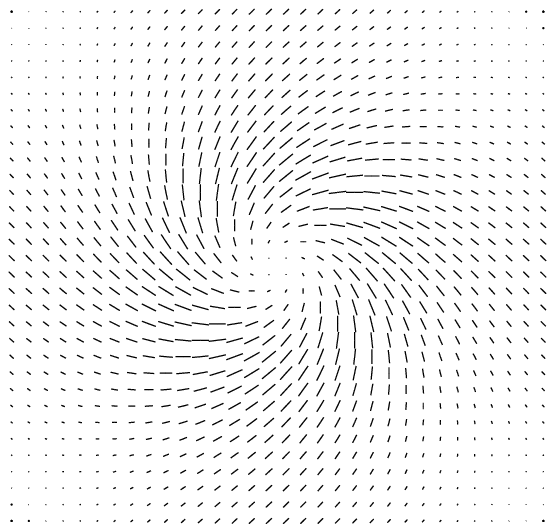}}
\caption{Gaussian ``hot spots.''  The left panel is purely
E-component, and the right panel is purely B-component.}
\label{fig:gauss}
\end{figure}

Detecting the gravity-wave signal in a CMB polarization data set
will be a daunting task: the tensor contribution to the CMB polarization
is expected to be considerably smaller than the contribution
due to scalar density perturbations.  Fortunately, the tensor
contribution has a geometrically different signature from the scalar
contribution, allowing the two to be separated.  To be specific,
any polarization map can be split into two components
\cite{ZalSel,KamKosSte}: the ``E component,''
which transforms as a scalar, and the ``B component,'' which transforms
as a pseudoscalar.\footnote{The division of a polarization
field into E and B components is closely analogous to the division
of a vector field into curl-free and divergence-free parts.
This is the reason for naming the components after the electric
and magnetic fields.  (Of course it is hardly necessary to
add that these names have nothing to do with the actual
electric and magnetic fields of the radiation.)  
Elsewhere in the literature,
the two components are called G and C, for ``gradient'' and ``curl'' respectively.
If the world needed yet another name for these components (which it
certainly doesn't), they could be called $+$ and $\times$;
as we will see below, these give the orientation of the
polarization with respect to the wavevector in Fourier space.
}
Scalar density perturbations produce only the E component,
leaving the B component as a clean probe of subdominant sources
of polarization such as tensor perturbations.\footnote{In addition to 
tensors, vector perturbations also produce
B-component polarization; however, because vector perturbations
decay over time, they
are not expected to contribute significantly to the observed
polarization, at least in the inflation-inspired models
we consider in this paper.  In some models (topological-defect
models in particular),
vector perturbations may be sourced at times close
to recombination and may therefore be observable, but we
do not consider such models in this paper.}

Given a full-sky polarization map, it is possible to separate the
E and B components perfectly; with incomplete sky coverage, however,
there is inevitably some cross-contamination between the components.
This naturally makes detecting the B-component more difficult, as it
is in danger of being swamped by the (typically much larger) E-component.
In this paper, we will examine the experimental requirements
to detect both E and B-component polarization signals in a degree-scale
experiment, accounting for this cross-contamination.

One approach to separating E from B is to observe in a 
circular ring \cite{Kea,Zal2,ChiMa}.  The separation
of components is particularly clean in this case, but
a strategy involving a two-dimensional map is
likely to be much better for measuring power spectrum
amplitudes \cite{TegCos}, as many more independent
modes at a given scale can be found in the data.
Attention has therefore been paid to 
finding normal modes that minimize the complications
due to E-B cross-contamination in
a two-dimensional map \cite{TegCos,LewChaTur}.  
In the present paper, we adopt
a more straightforward approach: we consider the likelihood
function of a polarization map in pixel space and compute
the Fisher matrix for the normalizations of both E and B power spectra.
Since this ``brute-force'' approach is based on the likelihood function
of the full data, it must be at least as good as (\textit{i.e.}, give as small\
error bars as) any method based on an expansion in normal modes.

The remainder of this paper is organized as follows.
In Section \ref{sec:eb}, we review some properties of polarization maps
and illustrate the difficulties in splitting a partial-sky map
into E and B components.
In Section \ref{sec:formalism}, 
we present the Fisher-matrix formalism we will use
to determine the detectability of the two components in a given experiment
and also present two simple approximation schemes for determining
the detectability.
Section \ref{sec:results} presents our results, and 
Section \ref{sec:disc} contains a brief discussion.

\section{The E-B Decomposition}
\label{sec:eb}

In this section we describe the nature of the E-B decomposition
of a polarization field.  This description makes no pretense
of completeness; much more information on this subject
can be found in the literature.  See in particular \cite{HuWhi,Zal}
and references therein.

\subsection{The flat-sky approximation}

Polarization is a spin-2 quantity (\textit{i.e.}, it is invariant under
180$^\circ$ rotations).  The linear polarization expected to be found
in the CMB is described by the two Stokes parameters $Q$ and $U$,
which are related to the magnitude $P$ and direction $\phi$
of the polarization as follows:
\begin{eqnarray}
Q &=&P\cos 2\phi,\\
U &=&P\sin 2\phi.
\end{eqnarray}

\begin{figure}[t!]
\centerline{\epsfxsize 2in\epsfbox{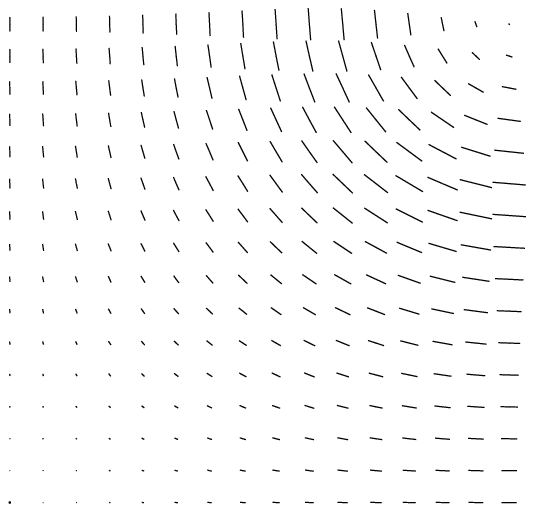}
\epsfxsize 2in\epsfbox{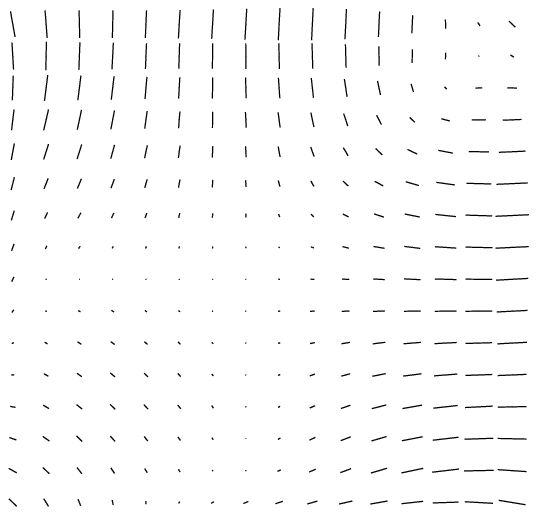}
\epsfxsize 2in\epsfbox{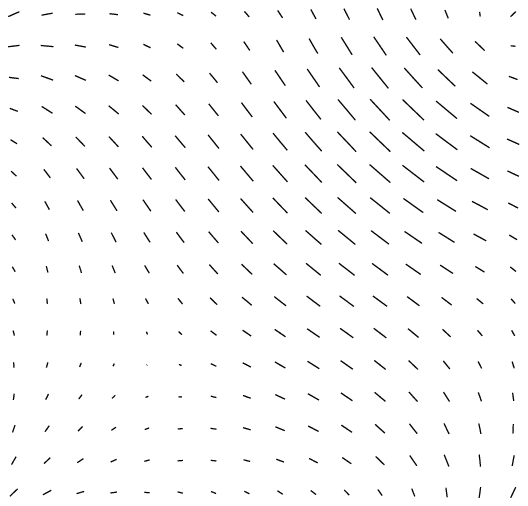}}
\caption{The left panel is simply one quadrant of the E-component
hot spot shown in Figure \ref{fig:gauss}.  The center
and right panels are a pure E and a pure B component respectively,
which sum to equal the left panel.}
\label{fig:gauss2}
\end{figure}

There are several ways to describe the division of a polarization
map into E and B components.  One simple way is to examine a
small patch of the sky, which can be well approximated as flat.
In this approximation,
an E-component polarization field is one whose Stokes parameters satisfy
the following equation:
\begin{equation}
2{\partial^2Q\over\partial x\partial y} -
\left({\partial^2U\over\partial x^2}-{\partial^2U\over\partial y^2}
\right)=0,\qquad\qquad\mbox{(E)}
\label{eq:econd}
\end{equation}
while a B-component field satisfies this equation:
\be
2{\partial^2U\over\partial x\partial y} +
\left({\partial^2Q\over\partial x^2}-{\partial^2Q\over\partial y^2}
\right)=0.\qquad\qquad\mbox{(B)}
\ee
(These equations are simply the spin-2 analogues of
the equations $\nabla\times\vec v=0$ and $\nabla\cdot\vec v=0$
for the scalar and pseudoscalar components of a vector field.)

An arbitrary polarization field can always be written
as a sum of an E and a B component.  One easy way to 
see this is to work in Fourier space.
If we consider a mode
with spatial dependence $e^{i{\mathbf k}\cdot{\mathbf x}}$, then
equation (\ref{eq:econd}) implies
that the E component has a polarization that is always parallel or
perpendicular to $\mathbf k$.
In terms of the
Stokes parameters, it can be written
\begin{equation}
\left(\matrix{Q\cr U}\right)_E \propto \left(\matrix{\cos 2\phi_{\mathbf k}\cr
\sin 2\phi_{\mathbf k}}\right)e^{i{\mathbf k}\cdot{\mathbf x}},
\label{eq:foure}
\end{equation}
where $\phi_{\mathbf k}$ is the angle $\mathbf k$ makes with the $x$ axis.
Similarly, a $B$-mode has a polarization direction that is always at
a 45$^\circ$ angle to $\mathbf k$.  It can be obtained by simply rotating
the polarization of an $E$ mode through 45$^\circ$ at each point:
\begin{equation}
\left(\matrix{Q\cr U}\right)_B \propto \left(\matrix{-\sin 2\phi_{\mathbf k}\cr
\cos 2\phi_{\mathbf k}}\right)e^{i{\mathbf k}\cdot{\mathbf x}}.
\label{eq:fourb}
\end{equation}

To decompose an arbitrary map into E and B components, therefore,
we simply take the Fourier transforms of $Q$ and $U$ and then,
for each wavevector, project $(Q,U)$ onto the axes defined
by equations (\ref{eq:foure}) and (\ref{eq:fourb}).
Of course, to perform this decomposition, we need to know
$Q$ and $U$ over all space.  If we only know $Q$ and $U$ over
a finite region, the Fourier transform and
hence the E-B split will depend on the boundary conditions we assume.

Figure \ref{fig:gauss} shows Gaussian ``hot spots'' for
both the E and B components.
Note that the B component, unlike
the E component, has handedness; this is the reason that density
perturbations, which lack handedness, cannot excite this component.
Of course, we could generate another pair of E, B maps by
replacing $(Q,U)$ by $(-Q,-U)$, which is equivalent
to rotating the polarizations in these plots through 90$^\circ$ at
each point.

To illustrate the ambiguity in the E-B decomposition of an incomplete
data set, suppose that we have observed only one quadrant of
the E-component hot spot, as shown in the left panel of
Figure \ref{fig:gauss2}.
This is of course a pure E polarization field: 
it satisfies equation (\ref{eq:econd})
at every point.  However, it is also the sum of the E and B
polarization fields shown in the center and right panels.
These fields were produced by Fourier transforming the region covered
by the data with periodic boundary conditions and splitting each
Fourier mode into E and B parts according to equations
(\ref{eq:foure}) and ({\ref{eq:fourb}}).  There are infinitely many
other ways the decomposition could have been done; for instance, the
data could have been padded with zeroes on all sides before Fourier
transforming.

In this particular example, the ``contamination'' of a pure E component
by B is not a small effect: the r.m.s.\ amplitude
of the B component shown in Figure \ref{fig:gauss2}
is 0.87 times that of the E component.
This is because the map being decomposed has a large amount of power
power on scales of order the width of the map.
In general, we can expect a large
amount of cross-contamination between E and B on the largest
scales probed by a given data set, with much less
contamination on smaller scales. 
The reason for this is simple:
the ambiguity in the E-B decomposition is a result of
our ignorance of the boundary conditions to impose on the
two components,
so the natural length scale associated with E-B mixing is the
width of the map.

\subsection{Beyond the flat-sky approximation}

Although the flat-sky formulae are in general simpler 
to work with, for many applications we need to consider
the exact, full-sky formulae.  In this section we will
very briefly summarize this formalism; see
\cite{ZalSel,KamKosSte,TegCos,LewChaTur} for further details
and useful identities.

Since polarization is a spin-2 quantity, the natural basis functions
to use in expressing the Stokes parameters $(Q,U)$ on the sphere are
the spin-2 spherical harmonics.  Specifically, we can write
\be
Q\pm iU = \sum_{l=2}^\infty \sum_{m=-l}^l a_{\mp 2,lm}\ {}_{\mp 2}Y_{lm},
\ee
where ${}_{\mp 2}Y_{lm}$ is a spin-2 spherical harmonic.
Detailed information on the spin-2 spherical harmonics can be found
in the sources
cited above.  For our purposes, all we need to know is that
decomposing a polarization field into E and B components is quite
simple in the spherical harmonic basis: the coefficients $a_{\mp 2,lm}$
in the expansion can simply be written
\be
a_{\pm 2,lm}=E_{lm}\pm iB_{lm}.
\ee
This, combined with the reality condition $a_{-2,lm}^*=(-1)^ma_{2,l-m}$,
allows one to determine $E_{lm}$ and $B_{lm}$ from the 
spherical harmonic coefficients.

If we assume that the polarization is a realization of a statistically
isotropic random process (\textit{i.e.}, that there is no preferred
direction), then the ensemble averages $\langle E_{lm}\rangle$
and $\langle B_{lm}\rangle$ must both vanish, and the 
covariances must satisfy
\begin{eqnarray}
\langle E_{lm}E_{l'm'}^*\rangle&=&C_l^E\delta_{ll'}\delta_{mm'},\\
\langle B_{lm}B_{l'm'}^*\rangle&=&C_l^B\delta_{ll'}\delta_{mm'}.
\end{eqnarray}
If in addition the random process is parity-invariant (lacks
handedness), then
\be
\langle E_{lm}B_{l'm'}^*\rangle = 0.
\ee
Furthermore, if the random process is Gaussian, then
the two power spectra $C_l^E$ and $C_l^B$ form a complete
description of the random process.\footnote{There is also
the cross-correlation between the E component and the temperature
anisotropy.  We choose to focus exclusively on polarization
data in the present paper, though, ignoring temperature data.}
These power spectra are therefore the only thing
a Gaussian theoretical model needs to predict about CMB polarization.
Fortunately, theoretical models are capable of computing predicted
power spectra with great precision and speed \cite{ZalSel,KamKosSte},
for instance using the publicly available CMBFAST software
\cite{cmbfast}.

Of course, actual observations always involve convolving
the true polarization field with the telescope beam.  As long
as the beam is azimuthally symmetric and purely co-polar, this results
in a simple replacement of $C_l^{E,B}$ with $C_l^{E,B}b_l^2$,
where $b_l$ is the Legendre transform of the beam pattern.
For a Gaussian beam of width $\sigma_\mathrm{beam}$ (FWHM $= \sigma_\mathrm{beam}\sqrt{8\ln 2}$),
we have 
\be
b_l^2=\exp(-l(l+1)\sigma_\mathrm{beam}^2).
\ee
Throughout this paper, we will use $C_l^{E,B}$ to denote the
beam-smoothed power spectra.

In the next section, we will consider the analysis of data
from a hypothetical CMB polarization experiment.  The key
ingredients in the analysis are the real-space correlations
between measurements at different points, 
$\langle Q({\mathbf x_1})Q({\mathbf x_2})\rangle$,
$\langle U({\mathbf x_1})U({\mathbf x_2})\rangle$, and 
$\langle Q({\mathbf x_1})U({\mathbf x_2})\rangle$,
which can be expressed as sums over the power spectra.
Specifically, if $Q$ and $U$ are defined with respect to coordinate
axes such that the $x$ axis joins the two points, then
\begin{eqnarray}
\langle Q({\mathbf x}_1)Q({\mathbf x}_2)\rangle
&=&
\sum_l\left(2l+1\over 4\pi\right)\left(C_l^EF_{1l}({\mathbf x}_1\cdot {\mathbf x}_2)
- C_l^BF_{2l}({\mathbf x}_1\cdot {\mathbf x}_2)\right),\\
\langle U({\mathbf x}_1)U({\mathbf x}_2)\rangle
&=&
\sum_l\left(2l+1\over 4\pi\right)\left(C_l^BF_{1l}({\mathbf x}_1\cdot {\mathbf x}_2)
- C_l^EF_{2l}({\mathbf x}_1\cdot {\mathbf x}_2)\right),\\
\langle Q({\mathbf x}_1)U({\mathbf x}_2)\rangle&=&0,
\end{eqnarray}
where the functions $F_{1l}$ and $F_{2l}$ can be expressed in terms
of Legendre functions as
\begin{eqnarray}
F_{1l}(x)&=&
2\left(
{(l+2)x\over (1-x^2)}P_{l-1}^2(x)-\left({l-4\over 1-x^2}+{l(l-1)\over 2}
\right)P_l^2(x)\over
(l-1)l(l+1)(l+2)\right),\\
F_{2l}(x) &=&
4\left(
(l+2)P_{l-1}^2(x)-(l-1)xP_l^2(x)\over
(l-1)l(l+1)(l+2)(1-x^2)\right).
\end{eqnarray}

In practice, we often wish to know the correlations
in some other coordinate system.  Since we know how
$(Q,U)$ transforms under rotations, we can easily
get these correlations by applying the appropriate 
rotation matrices to the correlations given
above.  (A pleasingly explicit recipe for doing this
can be found in \cite{TegCos}.)

\section{Formalism}
\label{sec:formalism}

\subsection{Likelihoods and Fisher Matrices}

\label{sec:fisher}

The Fisher information matrix provides a useful way to quantify
the ability of a data set to estimate parameters.  It has
been applied to great effect in the study of CMB temperature anisotropy
(\textit{e.g.}, \cite{TegTayHea}) and also
to CMB polarization studies \cite{TegCos,JKW}.
In this section, we show how the Fisher matrix can be used
to calculate the significance with which the amplitudes
of the E and B power spectra can be measured from a polarization map.

Suppose we have made maps containing $Q$ and $U$ measurements
at $N$ pixels.  The pixel locations are $\vec x_1,\ldots,\vec x_N$.
Our $2N$ data points can be written
\begin{eqnarray}
q_i&=&Q(\vec x_i)+\epsilon_{Qi},\\
u_i&=&U(\vec x_i)+\epsilon_{Ui}.
\end{eqnarray}
Here $i$ ranges over pixels in the map, and
$\epsilon_{Qi}$ ($\epsilon_{Ui}$) is the noise in the $i$th pixel
of the $Q$ ($U$) map.
We will assume uncorrelated noise:
\be
\langle\epsilon_{Ai}\epsilon_{Bj}\rangle=\sigma_{Ai}^2\delta_{AB}\delta_{ij}.
\label{eq:sigmaai}
\ee
Here $A$ and $B$ range over $\{Q,U\}$, and $\sigma_{Ai}^2$ is the
noise variance in pixel $i$ of map $A$.

We will arrange our data points into a data vector,
\be
\vec d=\left(\matrix{q_1 \cr \vdots \cr q_N \cr u_1 \cr \vdots \cr u_N}\right).
\ee
The $2N\times 2N$ data covariance matrix is defined to be
${\mathbf M}=\langle \vec d\vec d^T\rangle$.  
We can label elements of the data vector with a pair of indices $iA$,
with $i$ ranging from 1 to $N$ and $A$ being either $Q$ or $U$.
Then a typical element of the covariance matrix is $M_{iAjB}$.
The covariance matrix contains contributions from signal and noise
\be
M_{iAjB}=M^S_{iAjB}+M^N_{iAjB},
\ee
with
\begin{eqnarray}
M^S_{iQjQ}&=&w_Q(\vec x_i,\vec x_j)\equiv
\langle Q(\vec x_i)Q(\vec x_j)\rangle,\\
M^S_{iUjU}&=&w_U(\vec x_i,\vec x_j)\equiv
\langle U(\vec x_i)U(\vec x_j)\rangle,\\
M^S_{iQjU}&=&w_X(\vec x_i,\vec x_j)
\equiv\langle Q(\vec x_i)U(\vec x_j)\rangle
,\\
M^N_{iAjB}&=&\sigma^2_{iA}\delta_{ij}\delta_{AB}.
\end{eqnarray}
As described
in the previous section, 
the correlation functions $w_Q,w_U,w_X$ can be expressed as sums
over the power spectra.

A theory predicts a pair of power spectra $C^E_l$ and $C^B_l$
and hence a covariance matrix ${\mathbf M}$.
The likelihood of a theory is 
\be
{\cal L}({\mathbf M})\propto (\mbox{det }{\mathbf M})^{-1/2}\exp\left(-{1\over 2}
\vec d^T{\mathbf M}^{-1}\vec d\right).
\ee
It is more convenient to work with the quantity
\be
L=\ln\mbox{det } {\mathbf M}+\vec d^T{\mathbf M}^{-1}\vec d=
-2\ln {\cal L}+\mbox{const},
\ee
which can also be written in the following convenient form:
\be
L=\tr\left(\ln {\mathbf M}+{\mathbf M}^{-1}\vec d\vec d^T\right).
\ee
(The logarithm of a matrix is as usual defined via the Taylor
series, or equivalently by diagonalizing the matrix and taking the logarithms
of its eigenvalues.)

Now suppose that we are
considering a 
class of theories that contains
$P$ unknown parameters $\alpha_1\ldots,
\alpha_P$, and suppose for simplicity that the covariance matrix
is linear in these parameters, so that we can write
\be
{\mathbf M}={\mathbf M}^{(0)}+\sum_{p=1}^P(\alpha_p-1){\mathbf M}^{(p)}.
\ee
Here $\mathbf{M}^{(0)}$ represents the (unknown) true covariance matrix.
In other words, the true values of the parameters have been
taken to be one.  Our ability to measure parameters will
be determined by how sharply peaked the likelihood is about
its maximum, so we perform a Taylor expansion in $L$ to determine
this.
If we let $\partial_p$ stand for $\partial/\partial\alpha_p$, then
\be
\partial_pL=\tr\left({\mathbf M}^{-1}\partial_p{\mathbf M}-
{\mathbf M}^{-1}(\partial_p{\mathbf M}){\mathbf M}^{-1}\vec d\vec d^T\right)
=
\tr\left({\mathbf M}^{-1}{\mathbf M}^{(p)}-
{\mathbf M}^{-1}{\mathbf M}^{(p)}{\mathbf M}^{-1}\vec d\vec d^T\right).
\ee
Note that in the ensemble average, $\langle\vec d\vec d^T\rangle=
{\mathbf M}^{(0)}$,
so $\langle \partial_p L\rangle=0$ when all parameters are equal to one.

The quantity that characterizes the sharpness of the likelihood
peak is of course the second derivative, so we must plunge ahead
and differentiate again:
\be
\partial_q\partial_pL=
\tr\left(-{\mathbf M}^{-1}{\mathbf M}^{(q)}{\mathbf M}^{-1}{\mathbf M}^{(p)}+
{\mathbf M}^{-1}{\mathbf M}^{(q)}{\mathbf M}^{-1}{\mathbf M}^{(p)}{\mathbf M}^{-1}\vec d\vec d^T+
{\mathbf M}^{-1}{\mathbf M}^{(p)}{\mathbf M}^{-1}{\mathbf M}^{(q)}{\mathbf M}^{-1}\vec d\vec d^T
\right).
\ee
%
Let us take an ensemble average of this
quantity and evaluate it at ${\mathbf M}={\mathbf M}^{(0)}$ (which
is both the true value and the ensemble-average
maximum-likelihood location).  Then
\be
2F_{qp}\equiv\left<\left.\partial_q\partial_p L\right|_{{\mathbf M}={\mathbf M}^{(0)}}
\right>
=
\tr\left({{\mathbf M}^{(0)}}^{-1}{\mathbf M}^{(q)}{{\mathbf M}^{(0)}}^{-1}{\mathbf M}^{(p)}\right)
\label{eq:fpq}
\ee
The quantities $F_{qp}$ are the elements of the $P\times
P$ Fisher matrix ${\mathbf F}$.
They tell us the expected uncertainty with which a parameter
$\alpha_p$ can be determined from a likelihood analysis.
Specifically, if all parameters except the $p$th
are known \textit{a priori}, then $\alpha_p$ will be
determined with an expected error of
$1/\sqrt{F_{pp}}$.  If on the other
hand all parameters are unknown, then the uncertainty
is in $\alpha_p$ is $\sqrt{({\mathbf F}^{-1})_{pp}}$.
Furthermore, these expected uncertainties
are the smallest that can be obtained from this
data set by any unbiased data analysis method.

We will be interested in determining whether the E- and B-component
polarizations can be detected by a given experiment.  Let us
make an admittedly optimistic assumption: suppose that
the \textit{shapes} of the power spectra $C_l^E$ and $C_l^B$ are known
but the amplitudes are not.  Then we will have two parameters
$\alpha_E$ and $\alpha_B$, such that
\be
C_l^{E,B}=\alpha_{E,B}\hat C_l^{E,B},
\ee
where $\hat C_l^{E,B}$ represents the true power spectrum.
Then the matrices ${\mathbf M}^{(E)}$ and ${\mathbf M}^{(B)}$ that appear in equation
(\ref{eq:fpq}) are simply the parts of ${\mathbf M}$ (and 
in particular ${\mathbf M}^{S}$) proportional to $C_l^{E}$ and $C_l^{B}$.

We will say that the E (B) component is detectable if
the parameter $\alpha_E$ ($\alpha_B$) can be measured with an expected
uncertainty considerably less than one.  Specifically, if $\alpha_E$
has an expected uncertainty of $1\over q$, then we can
expect to measure the amplitude of $C_l^E$ with a signal-to-noise
ratio of $q$.
So if, say, we are interested in knowing whether a particular
experimental design will provide a 3-sigma detection
of the E component, we simply 
compute the
$2\times 2$ Fisher
matrix ${\mathbf F}$ and determine whether $\sqrt{({\mathbf F}^{-1})_{EE}}>3$.

\subsection{The JKW Approximation}

In the next section, we will present experimental requirements for
detecting the E and B components, using the formalism described above.
This process is somewhat laborious, since it involves manipulating
$2N\times 2N$ matrices, so simpler
methods are
clearly desirable.  One very useful
such approximation has been provided by
Jaffe, Kamionkowski, \& Wang (hereinafter 
JKW) \cite{JKW}.  Similar results may be found in \cite{Zal2}.
In this section, we present
a brief heuristic ``derivation'' of the JKW approximation.

First, consider an experiment in which the $N$ pixels
cover the entire sky uniformly.  Let us also suppose
that the noise level $\sigma^2$
is the same in each pixel of the $Q$ and $U$ maps.  (In other words, $\sigma^2$
is the same as $\sigma_{Ai}^2$ of equation (\ref{eq:sigmaai}) and
is assumed to be the same across all pixels of both maps.)
The analysis of this experiment is quite simple: we can
estimate each coefficient $a_{\pm 2,lm}$, and
hence each $E_{lm}$ and $B_{lm}$, independently
by exploiting the orthonormality of the spherical harmonics
over the whole sphere.  Each coefficient will have
noise of amplitude
\be
\sigma_*^2={4\pi\over N}\sigma^2.
\label{eq:sigmastar}
\ee
It is convenient to define 
the weight $w$ of a set of $Q$ and $U$ maps to be\footnote{Beware:
this definition differs by a factor $4\pi$ from the weight
as defined in JKW.  While we're on the subject, note that
the polarization power spectra
in JKW and also in Ref.
\cite{KamKosSte} 
differ by a factor of two
from those found elsewhere in the literature and in this paper.}
\be
w\equiv\sum_{i=1}^{N}\sum_{A=\{Q,U\}}\sigma_{Ai}^{-2}.
\ee
The results of this section depend on the assumption of uniform noise,
so $w=2N/\sigma^2$.  Equation (\ref{eq:sigmastar}) can therefore
be written
\be
\sigma_*^2={8\pi\over w}.
\ee
If we wish to estimate the E-component power spectrum $C_l^E$,
therefore, we have at our disposal a set of independent Gaussian
random numbers $\hat E_{lm}$ (estimators of the true coefficients
$E_{lm}$) with zero mean and variances 
\be
\langle|\hat E_{lm}|^2\rangle = C_l^E+\sigma_*^2.
\ee
Since all the variables are independent, the likelihood has
the usual Gaussian form:
\be
{\cal L}\propto \left(\prod_{l,m}(C_l^E+\sigma_*^2)\right)^{-1/2}
\exp\left(-{1\over 2}\sum_{l,m}{|E_{lm}|^2\over
C_l^E+\sigma_*^2}\right).
\ee

If we assume as before that only the overall normalization
of the power spectrum is unknown, we can determine
its expected uncertainty just as in the previous
section, by computing the Fisher matrix (which
is diagonal for a full-sky experiment).
Differentiating the above expression for $\cal L$ twice,
we find that
the uncertainty in the parameter $\alpha_E$ is given by
\be
(SNR)_E^2\equiv
\sigma_{\alpha_E}^{-2}={1\over 2}\sum_{l,m} \left(C^E_l\over C^E_l+\sigma_*^2
\right)^2
={1\over 2}\sum_{l=2}^\infty {2l+1
\over(1+(8\pi/wC_l^E))^2}.
\label{eq:jkwfullsky}
\ee
Here SNR stands for ``signal-to-noise ratio.''
An experiment with an SNR of, say, 3, is one in which we
would expect to measure the power spectrum normalization
with an accuracy of three sigma.
Of course, an identical relation applies to $\alpha_B$.

This result has a very simple interpretation \cite{Zal2}.  The quantity
$wC_l^E/8\pi$ is the square of the signal-to-noise ratio for a single mode
at multipole $l$.  Modes for which this quantity is much greater
than unity contribute 1 to the sum, while modes for which
it is much less than one contribute nothing.  Therefore,
what this formula says, roughly, is that the
square of the signal-to-noise ratio of the power spectrum
amplitude is equal to $1\over 2$ the number of modes
that are detected with high signal-to-noise.

Now consider an experiment that only covers a fraction
of the sky $f_{\rm sky}$.  The number of independent
modes that can be detected will of course be reduced
by a factor $f_{\rm sky}$.  On the other hand, the
noise variance for each mode will be reduced by the same
factor, since the total weight of the experiment is concentrated
in a smaller area.  We might guess, therefore, that
the generalization of equation (\ref{eq:jkwfullsky}) to
the case of partial sky coverage is
\be
(SNR)_{E,B}^2={f_{\rm sky}\over 2}\sum_{l}{2l+1
\over(1+(8\pi f_{\rm sky}/wC_l^{E,B}))^2}.
\label{eq:jkw}
\ee
Equation (\ref{eq:jkw}) is the JKW approximation (compare to equation
(1) of JKW).

A question arises as to the lower limit of the sum.
JKW advocate starting the sum at $l_{\rm min}=180^\circ/L$,
where $L$ is the survey size, on the grounds that modes with smaller
$l$ cannot be probed.  It can be argued, though,
that the sum should begin at the lowest possible $l$ (namely 2).
After all, the fact that modes with low $l$ are not probed
is already accounted for by the inclusion of the prefactor
$f_{\rm sky}$.   The effective number of independent modes below some
multipole $l_0$ is approximately $f_{\rm sky}l_0^2$, which is
better approximated by $f_{\rm sky}\sum_2^{l_0}(2l+1)$ 
than by $f_{\rm sky}\sum_{l_{\rm min}}^{l_0}
(2l+1)$.  The choice made makes little difference to the final results
(at most about 20\%, and usually much less, for the results to be shown below),
so
for consistency we follow JKW's prescription.

Of course, the JKW approximation does not account for the
mixing of E and B modes due to incomplete sky coverage.
We therefore expect it to overestimate the detectability
of the subdominant B component, since some modes that are used
to estimate B will be swamped by contamination from the larger
E component.  As we will see below, this is indeed the case.
Perhaps more surprising is the fact that the JKW approximation
sometimes
\textit{underestimates} the detectability of the E component.  
In the next subsection, we will consider an enhanced version
of the JKW approximation that sheds some light on the reason
for this.

\subsection{A toy model of E-B mixing}
In this section, we will consider an experiment
that covers a square patch of sky that is small
enough to permit the use of the flat-sky approximation.
Suppose as usual that the sky polarization
is a Gaussian random field ${\mathbf p}=\left(\matrix{Q\cr U}\right)$.
It can of course be written as a Fourier transform,
\be
{\mathbf p}=\int d^2k\,\tilde{\mathbf p}({\mathbf k})e^{i{\mathbf k}\cdot {\mathbf x}},
\ee
and each Fourier component can be split into an E and a B part:
\be
\tilde{\mathbf p}({\mathbf k})=\tilde E({\mathbf k}) \left(\matrix{\cos 2\phi_{\mathbf k}\cr
\sin 2\phi_{\mathbf k}}\right)
+\tilde B({\mathbf k}) \left(\matrix{-\sin 2\phi_{\mathbf k}\cr \cos 2\phi_{\mathbf k}}\right).
\ee

The assumption that the polarization is a realization
of a homogeneous, isotropic,
parity-invariant
Gaussian random field means that $\tilde E$ and $\tilde B$
are Gaussian random variables with zero mean and covariances
\begin{eqnarray}
\langle \tilde E({\mathbf k})\tilde E^*({\mathbf k'})\rangle&=& {1\over (2\pi)^2}
C^E(k)\delta({\mathbf k}-{\mathbf k'}),\\
\langle \tilde B({\mathbf k})\tilde B^*({\mathbf k'})\rangle&=& {1\over (2\pi)^2}
C^B(k)\delta({\mathbf k}-{\mathbf k'}),\\
\langle\tilde E({\mathbf k})\tilde B^*({\mathbf k'})\rangle &=&0.
\end{eqnarray}
The power spectra $C^E$ and $C^B$ depend only on the magnitude of
$\mathbf k$.  The factor of $(2\pi)^2$ is inserted for consistency
with standard normalization conventions: $C^{E,B}(k)\approx C^{E,B}_l$
with $l=k$.

Now, suppose that the polarization is measured over a 
square patch of sky of area $L^2$.  
We might\footnote{In fact, we probably would not
analyze the data in quite such a na\"\i ve way; we would at least
taper the edges to reduce ringing in the Fourier modes.
But we're trying to keep things simple.}
 choose to analyze such a data set by 
expanding the observed region in a Fourier series with coefficients
\be
\tilde {\mathbf a}({\mathbf q})={1\over L^2}\int_{L^2}d^2x\,
{\mathbf p}({\mathbf x})e^{-i{\mathbf q}\cdot{\mathbf x}},
\ee
where ${\mathbf q}=(2\pi/L)(n_x,n_y)$ for integers $n_x$ and $n_y$.

These Fourier series coefficients are related to the true Fourier
transform in the usual way,
\be
\tilde {\mathbf a}({\mathbf q}) = \int d^2k\, \tilde{\mathbf p}({\mathbf k})W({\mathbf k}-
{\mathbf q}),
\ee
where the window function is
\be
W({\bm\gamma})=\left(\sin \gamma_xL/2\over \gamma_xL/2\right)
\left(\sin \gamma_yL/2\over \gamma_yL/2\right).
\ee
In other words, each mode $\tilde {\mathbf a}({\mathbf q})$ probes a range
of $\mathbf k$ values of width $\sim L^{-1}$ around $\mathbf q$.

Suppose that we wish to estimate the E and B power spectra
from these Fourier coefficients.  We might proceed by decomposing
each $\tilde {\mathbf a}({\mathbf q})$ into an ``E'' and ``B'' piece:
\be
\tilde {\mathbf a}({\mathbf q})=\tilde a_E\left(\matrix{\cos 2\phi_{\mathbf q}\cr
\sin 2\phi_{\mathbf q}}\right)+
\tilde a_B\left(\matrix{-\sin 2\phi_{\mathbf q}\cr
\cos 2\phi_{\mathbf q}}\right)
\ee
However, since $\tilde {\mathbf a}({\mathbf q})$ contains contributions
from a wide range of $\mathbf k$'s, not all of which are parallel
to $\mathbf q$, these will not really be purely E or B.  In fact,
the mean-square value of the supposedly E component is
\be
\langle |\tilde a_E({\mathbf q})|^2\rangle
=
{1\over (2\pi)^2}\int d^2k\,(C^E(k) \cos^2 2\alpha
+C^B(k) \sin^2 2\alpha) W^2({\mathbf k}-{\mathbf q}),
\ee
where $\cos\alpha=\hat{\mathbf k}\cdot\hat{\mathbf q}$.
A similar equation holds for the B component.

When $q$ is large compared to $1/L$, of course, all values of $\mathbf k$ that
contribute significantly to the integral are quite close in
direction to $\mathbf q$, so $\sin^22\alpha\approx 0$
and $\tilde a_E({\mathbf q})$ really does depend almost
entirely on the E component.  In other words, mixing of E
and B is relatively unimportant on small scales.
But for small $\mathbf q$, $\alpha$ cannot be taken to be small.
In fact, if we assume that the power spectra are approximately
constant over the range where the window function is large,
we can pull them out of the integral
to find
\be
\langle|\tilde a_E({\mathbf q})|^2\rangle\approx
{1\over L^2}\left(C^E(q)\overline{c^2_\mathbf{q}}+C^B(q)\overline{s^2_\mathbf{q}}\right),
\ee
where $\overline{c^2_\mathbf{q}}$ and $\overline{s^2_\mathbf{q}}$ 
are the averages
of $\cos^22\alpha$ and $\sin^22\alpha$ weighted by $W^2(\mathbf{k}-
\mathbf{q})$.
Similarly,
\be
\langle|\tilde a_B({\mathbf q})|^2\rangle\approx
{1\over L^2}
\left(C^B(q)\overline{c^2_\mathbf{q}}+C^E(q)\overline{s^2_\mathbf{q}}\right).
\label{eq:atildeb}
\ee
We can find these
averages by numerical integration.
Averaging over the direction  of
$\mathbf q$, we get approximately
\be
1-\overline{c^2_q}=\overline{s^2_q}={2.4\over qL}.
\ee
(If we are a bit
more sophisticated and taper the edges of the data before Fourier
transforming, then the window function $W$ doesn't ring so much,
and $\overline{s_q^2}$ can be reduced by $\sim10$-$20\%$.  As we will
see in the next section, though, this approximation gives surprisingly
good results as is.)

Now suppose that $C^E\gg C^B$.  Then for each $\mathbf q$ with a reasonably
large value of $\overline{s_q^2}$ we will have two 
independent\footnote{It can be shown that correlations between
the different modes are relatively weak.  This is not
terribly precise, but after all this section is called ``a toy model''!}
modes that are dominated by E, rather than the expected one:
according to equation (\ref{eq:atildeb}),
even the nominal B mode $\tilde a_B({\bf q})$ is mostly E!
Since, as we have seen in the previous section, the
signal-to-noise ratio is determined by counting the number
of modes with high signal-to-noise, this enhances
the significance with which $\alpha_E$ can be measured.

In fact, we can modify the JKW approximation to take
this into account.  Each ``nominal E mode'' $\tilde a_E(\mathbf{q})$
provides a measurement of the E-mode with mean-square amplitude
$C_l^E\overline{c_l^2}$ and mean-square noise $8\pi f_{\rm
sky}/w+C_l^B\overline{s_l^2}$.  (Here $|\mathbf{q}|=l$.
We are imagining an attempt to measure the E-component, so we
treat the B-component part of the signal as if it were noise.)
We can therefore replace 
$8\pi f_{\rm sky}/wC_l^E$ in equation (\ref{eq:jkw}) with
\be
\kappa_{1l}\equiv{8\pi f_{\rm sky}/ w+C_l^B\overline{s_l^2}\over
C_l^E\overline{c_l^2}}.
\ee
And we should also include a term that accounts for the fact
that $\tilde a_B({\bf q})$ can be used to measure the E-mode amplitude:
\be
\kappa_{2l}\equiv {8\pi f_{\rm sky}/ w+C_l^B\overline{c_l^2}\over
C_l^E\overline{s_l^2}}.
\ee
The final result is
\be
(SNR)_E^2={f_{\rm sky}\over 2}\sum_l (2l+1)
\left(
{1\over(1+\kappa_{1l})^2}
+
{1\over(1+\kappa_{2l})^2}\right)
.
\label{eq:jkwenh}
\ee

\begin{figure}[t!]
\centerline{\epsfxsize 4in\epsfbox{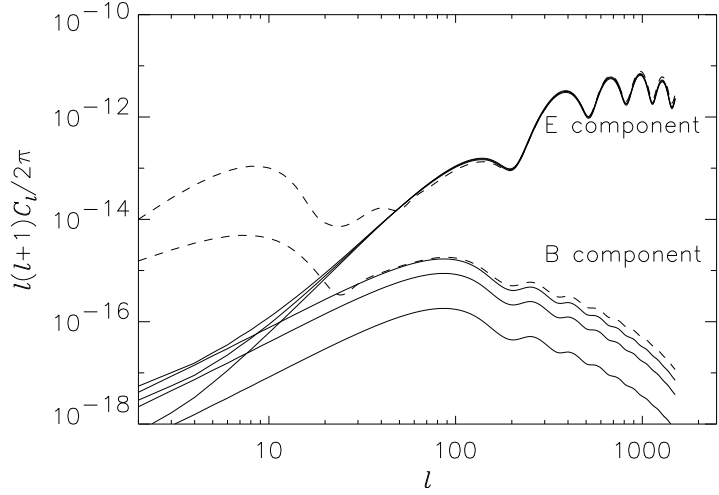}}
\caption{E and B Power Spectra.  The solid curves are the $\Lambda$CDM
models with no reionization described in the text.  From top to bottom,
$T/S$ takes the values 0.1, 0.05, 0.01.  The dashed curve is a reionized
model with $\tau=0.3$, $T/S=0.1$, and $n$ tilted to $1.15$.}
\label{fig:specs}
\end{figure}

Although this looks quite messy, the interpretation is
fairly simple.  $\kappa_{1l}^{-1}$ and $\kappa_{2l}^{-1}$ 
are the squared signal-to-noise ratios with which the E-component
amplitude can be detected in each mode (counting
the B-component contribution
as noise).  Modes with high
signal-to-noise contribute one to the sum, while modes with
low signal-to-noise (or low E-signal-to-B-signal, since B-signal
is being counted as noise) contribute zero.  In the no-mixing
limit, $\overline{s_l^2}\to 0$, the $\kappa_{2l}$ term doesn't
contribute, and the $\kappa_{1l}$ term agrees with equation (\ref{eq:jkw}).

Incidentally, we now face the same question as in the original
JKW approximation regarding the lower limit of the sum.  As before,
it makes relatively little difference whether we start
the sum at 2 or at $180^\circ/L$.  In the results shown below,
we have started the sum at 2.

We can try to use this approximation to
find the B-component detectability, but it turns out to give terrible
results.  This is not surprising: because we can always
measure the E component more accurately than the B component,
modeling it simply as an unknown source of noise
is a poor approximation.

\section{Results}
\label{sec:results}

\begin{figure}[t!]
\centerline{
\epsfxsize 3in\epsfbox{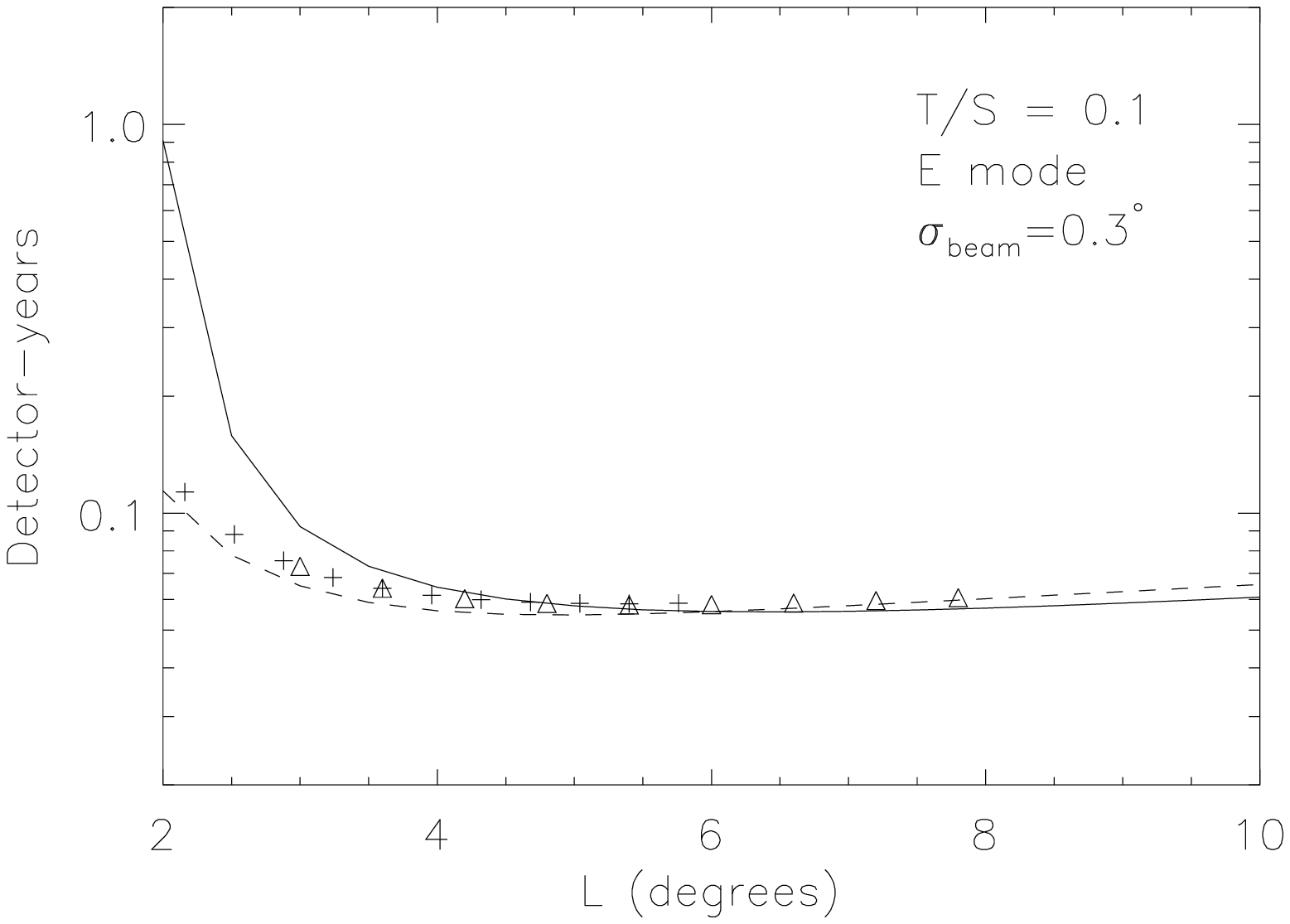}
\epsfxsize 3in\epsfbox{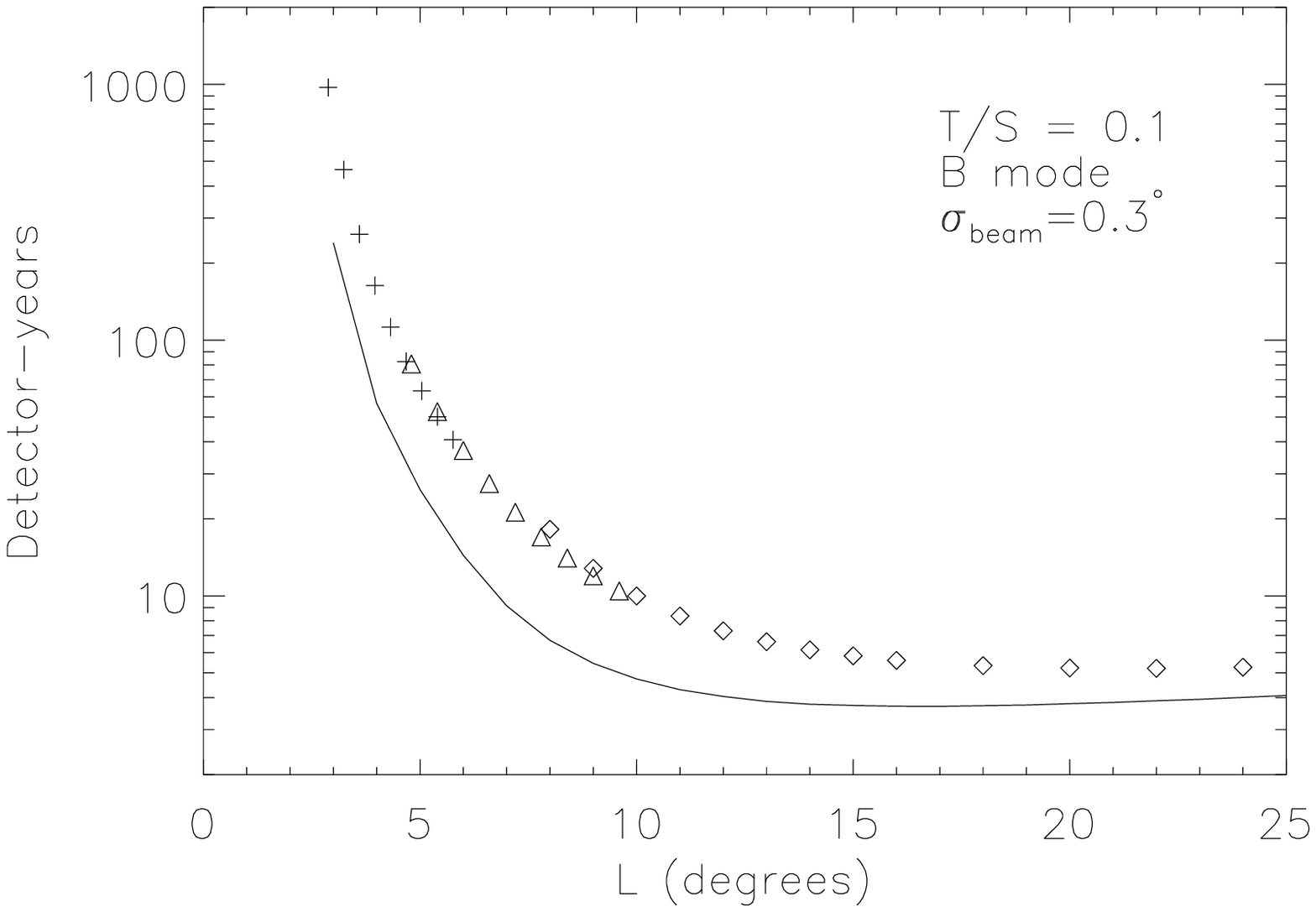}
}
\centerline{
\epsfxsize 3in\epsfbox{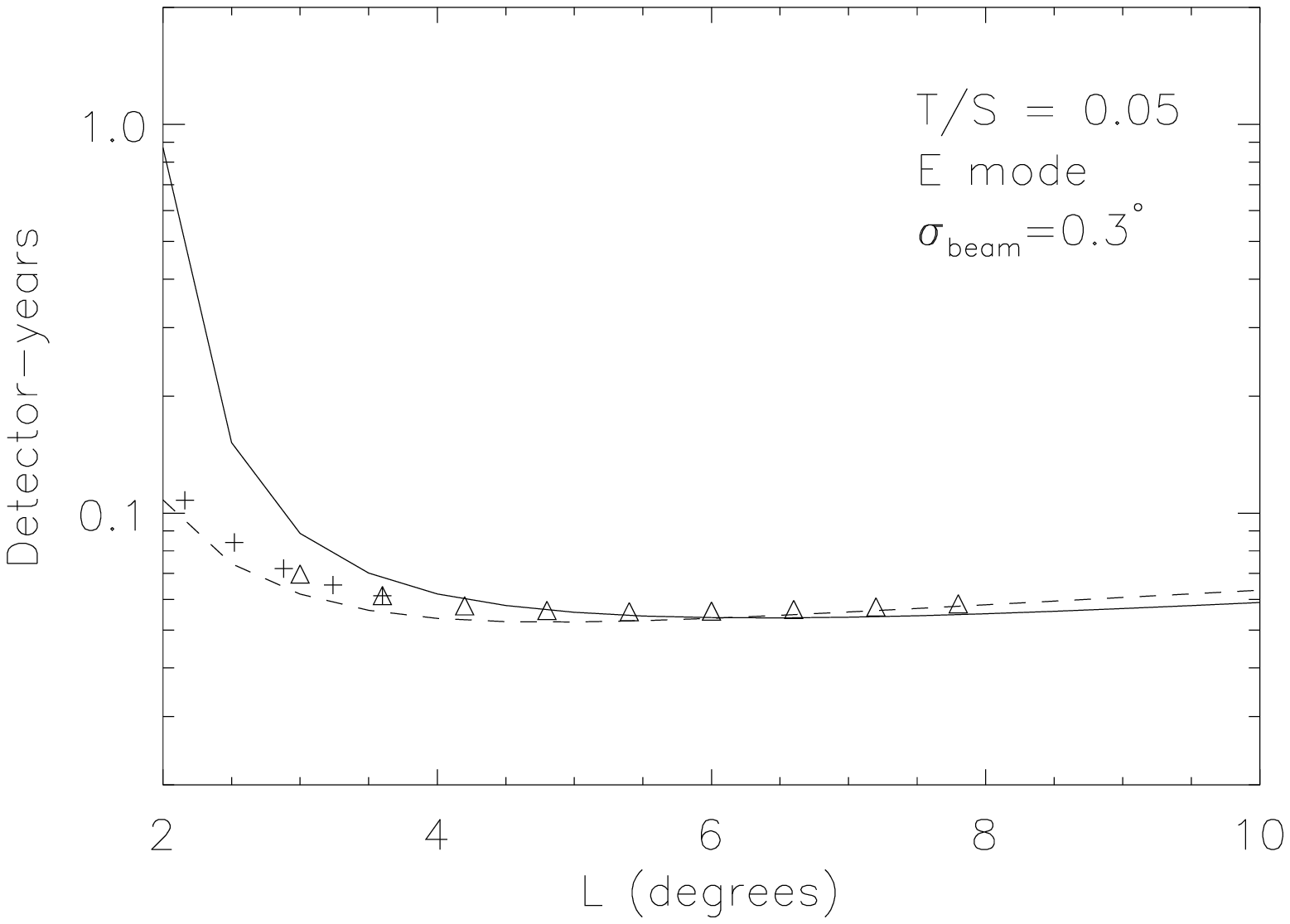}
\epsfxsize 3in\epsfbox{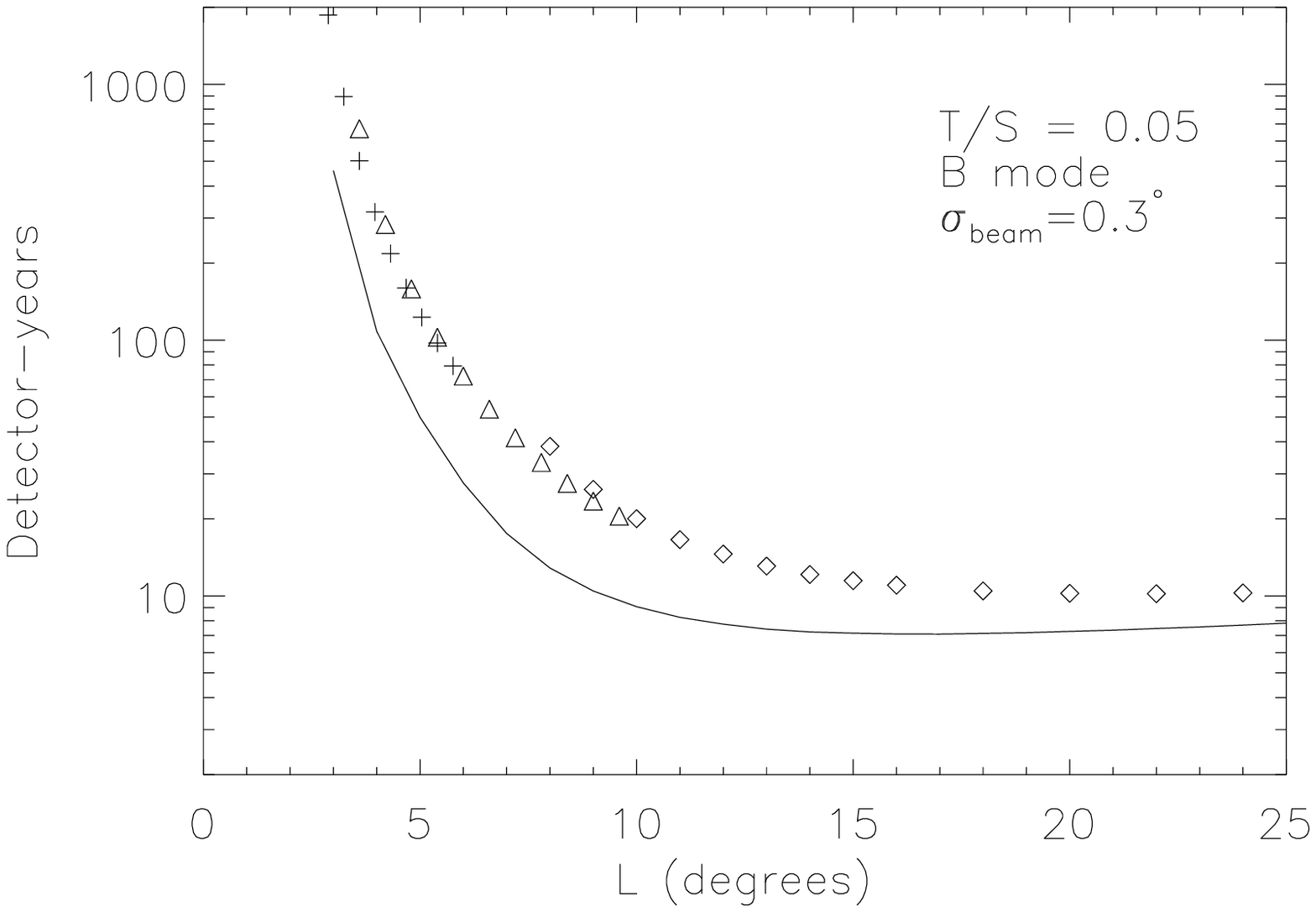}
}
\centerline{
\epsfxsize 3in\epsfbox{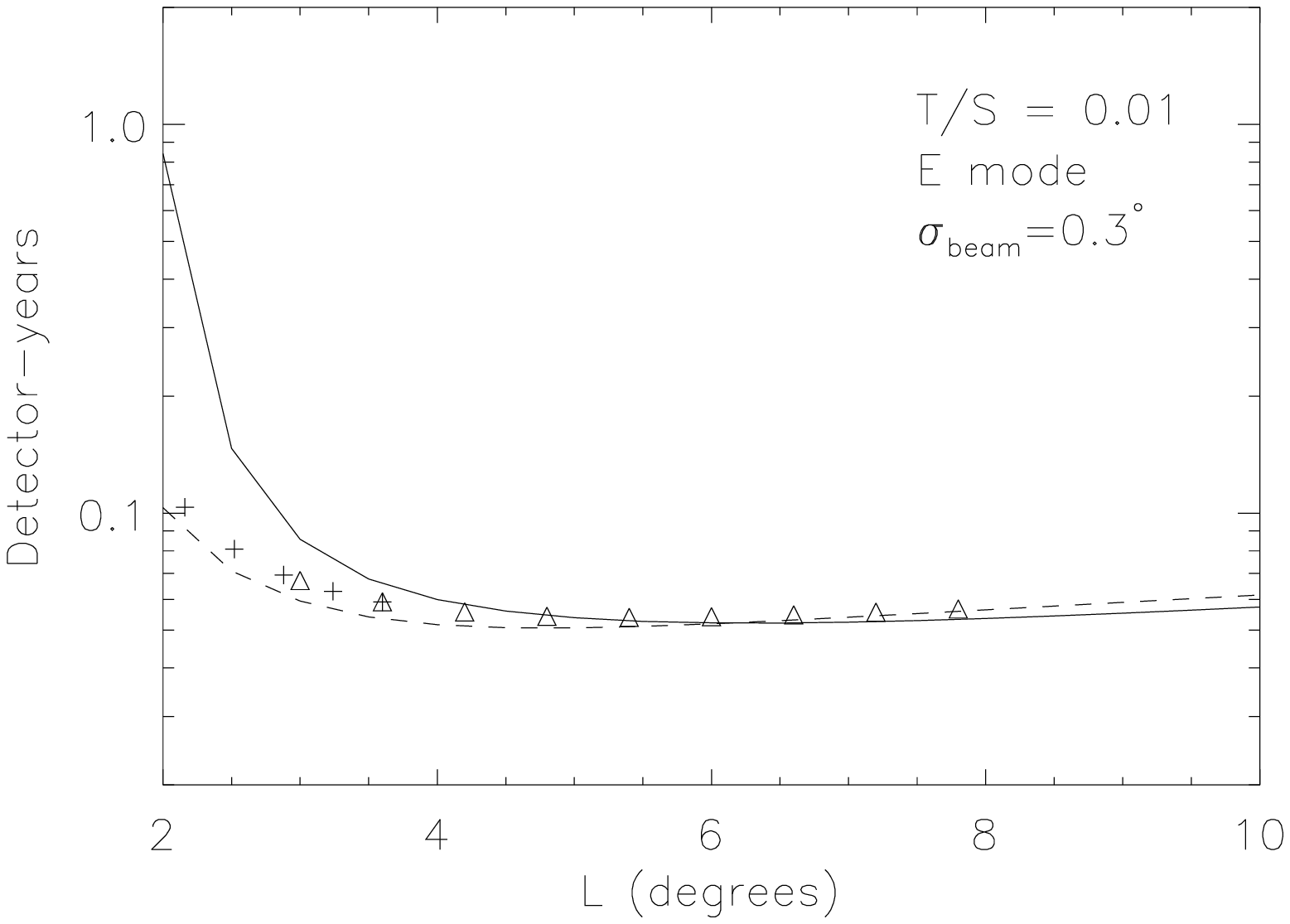}
\epsfxsize 3in\epsfbox{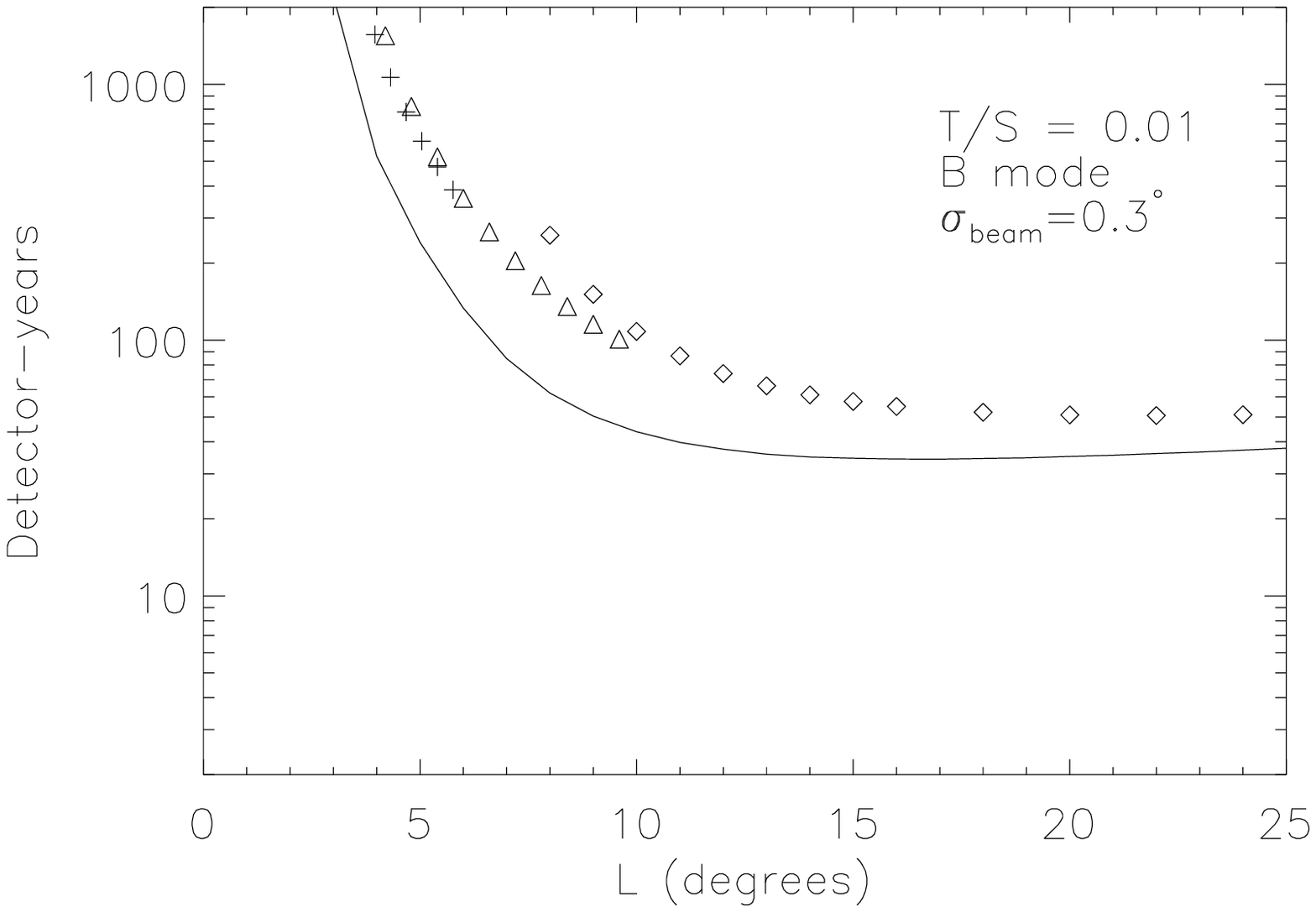}
}
\caption{E and B component detectability for different
tensor-scalar ratios.  In this figure and the next,
the vertical axis is the observing time in detector-years,
at a nominal sensitivity of 100 $\mu\rm K\,s^{1/2}$,
required for a 3-sigma measurement of the power spectrum
amplitude.
The beam size is $\sigma_{\rm beam}=0.3^\circ$ (FWHM=0.71$^\circ$).
The points are the results of the Fisher matrix analysis.  The
three different symbols are for maps with pixel sizes of $0.18^\circ$,
$0.3^\circ$, and $0.5^\circ$.  The solid line is the result
found with the JKW approximation.  In the E-component case, the
result of the enhanced JKW approximation is shown as a dashed line.
}
\label{fig:result1}
\end{figure}

Consider an experiment in which a square patch of sky
of length $L$ is observed.  We will suppose that 
the patch is pixelized into an $N_{\rm side}\times N_{\rm side}$ square array
and that both $Q$ and $U$ are measured in each pixel,
so that the dimension of our data vector $\mathbf d$ is $2N_{\rm side}^2$.
Suppose furthermore that the noise level $\sigma$ is
the same for all of these measurements.
The weight of the experiment is then
\be
w={2N_{\rm side}^2\over\sigma^2}.
\ee
The weight is of course proportional to the observing time:
$w={N_{\rm d}t_{\rm obs}T_0^2/s^2}$, where $N_{\rm d}$ is the number
of detectors and $t_{\rm obs}$ is the observing time.  $T_0=2.728$ K
is the current CMB temperature.  Its presence is simply to convert
units: we choose to measure the sensitivity $s$ in temperature units
but the power spectra in dimensionless $\Delta T/T$ units.
Instead of quoting
weights, therefore, we can simply quote observing times in, say,
detector-years at some nominal sensitivity level.  In the results
below, we will use $s=100 \,\mu\rm K\,s^{1/2}$ as our nominal sensitivity,
so our observing times in detector-years will simply be
\be
\mbox{Detector-years}\equiv
N_{\rm d}\left(t_{\rm obs}\over 1\,\mbox{year}\right)\left(100\,\mu{\rm K}
\,{\rm s}^{1/2}\over s
\right)^2
={w\over 2.35\times 10^{16}}.
\ee

We consider a ``concordance'' $\Lambda$CDM cosmological model with
parameters chosen for reasonable agreement with the bulk of the
available data: 
$h=0.70,\Omega_{\rm b}=0.041,\Omega_\Lambda=0.7,\Omega_{\rm tot}=1,
n=1$.  In most of our results, we ignore the effects of reionization by setting
the optical depth $\tau$ to last scattering equal to zero.
We allow the tensor-to-scalar ratio $T/S$, defined as the ratio
of the tensor and scalar temperature power spectra at $l=2$, to vary.
E and B power spectra for these models are shown in Figure \ref{fig:specs}.

\begin{figure}[t!]
\centerline{\epsfxsize 3in\epsfbox{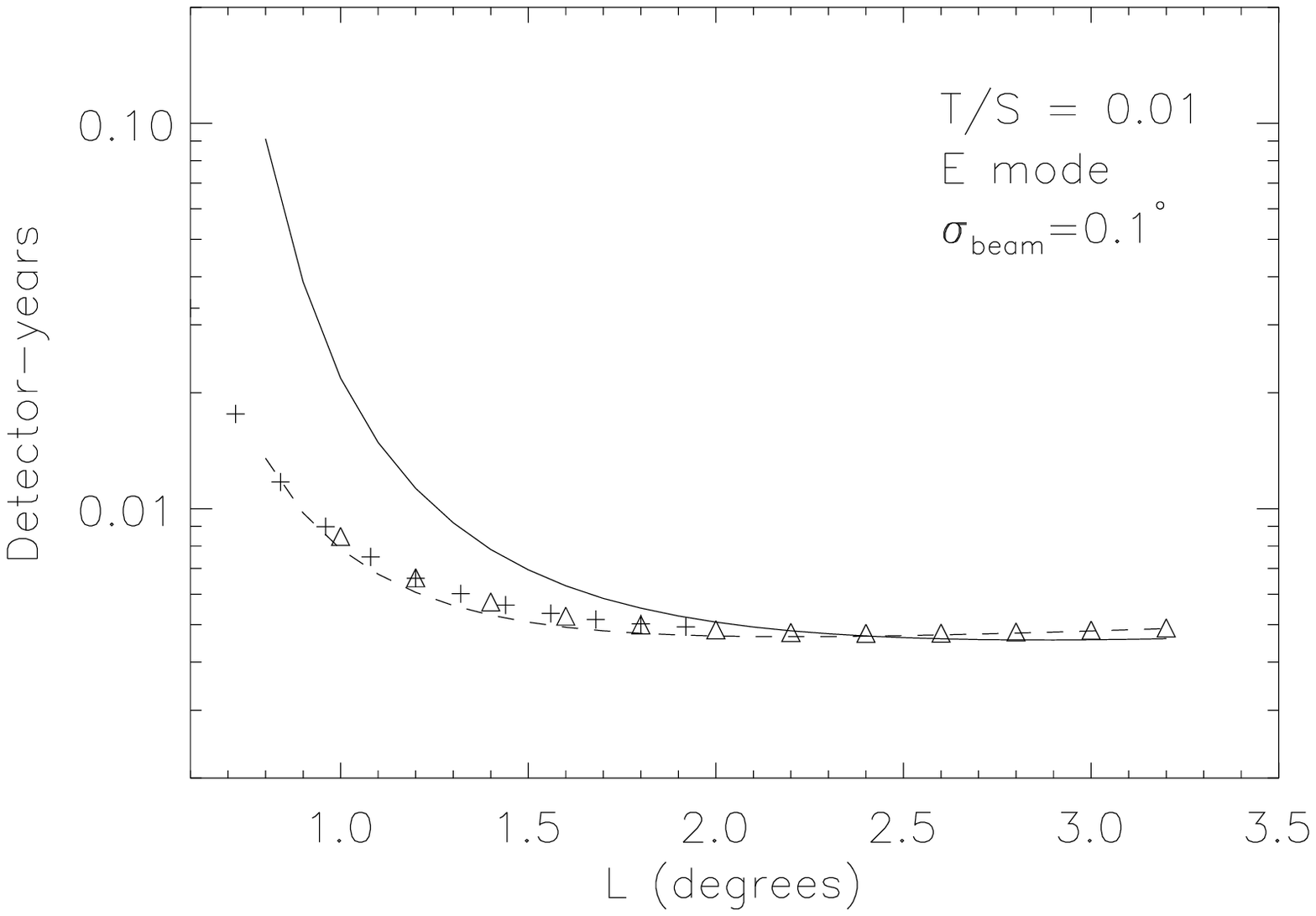}
\epsfxsize 3in\epsfbox{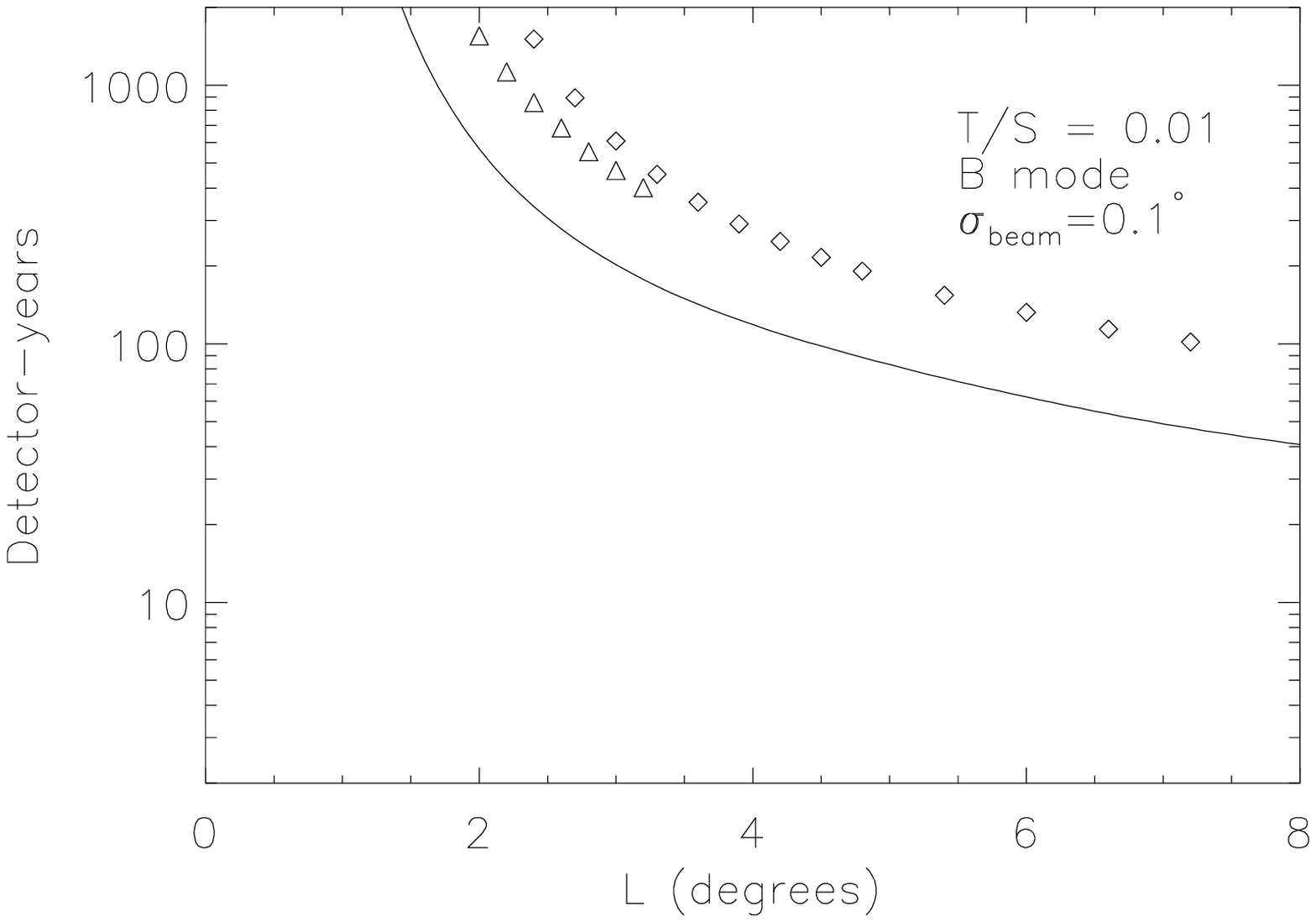}}
\centerline{\epsfxsize 3in\epsfbox{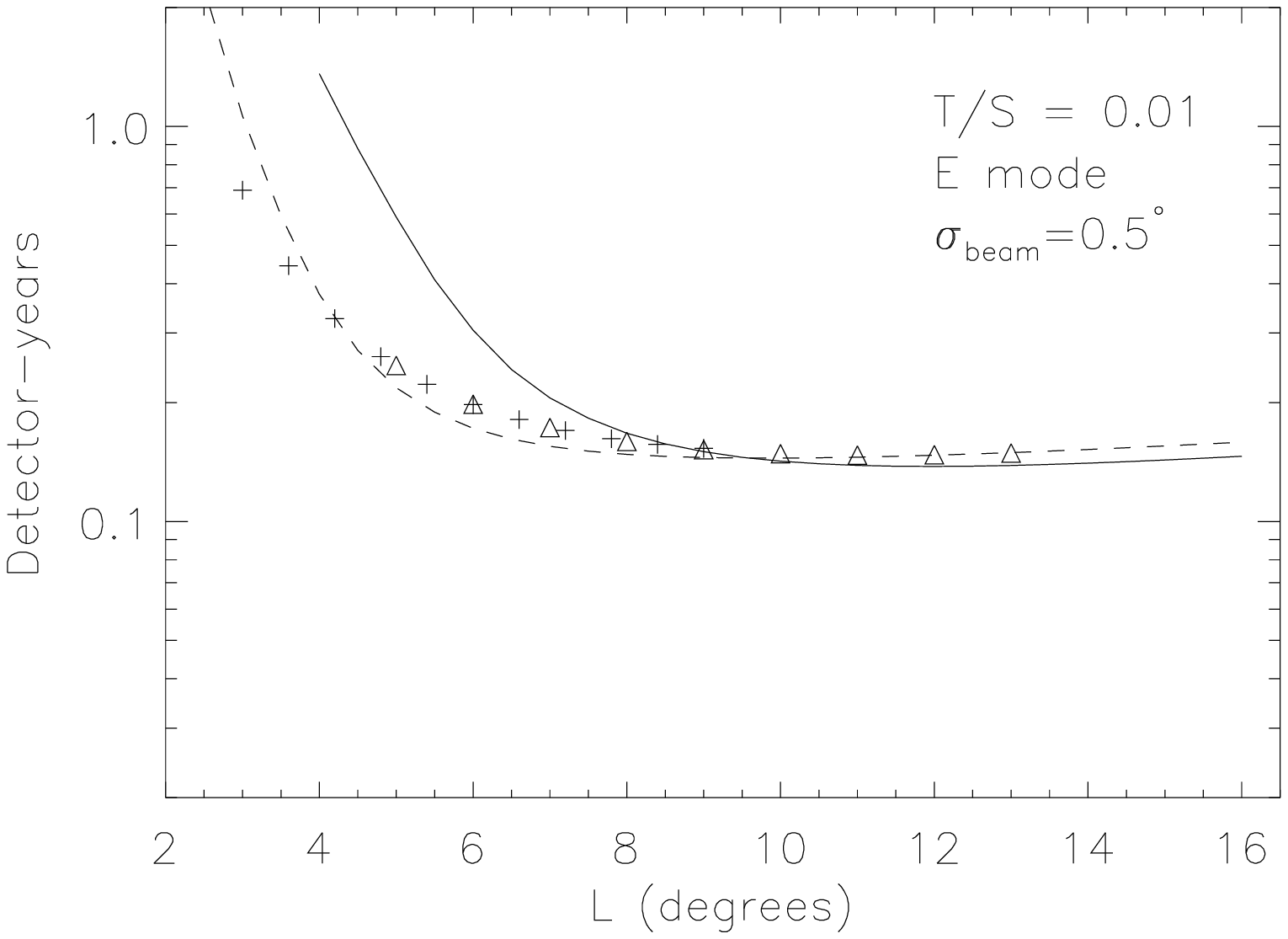}
\epsfxsize 3in\epsfbox{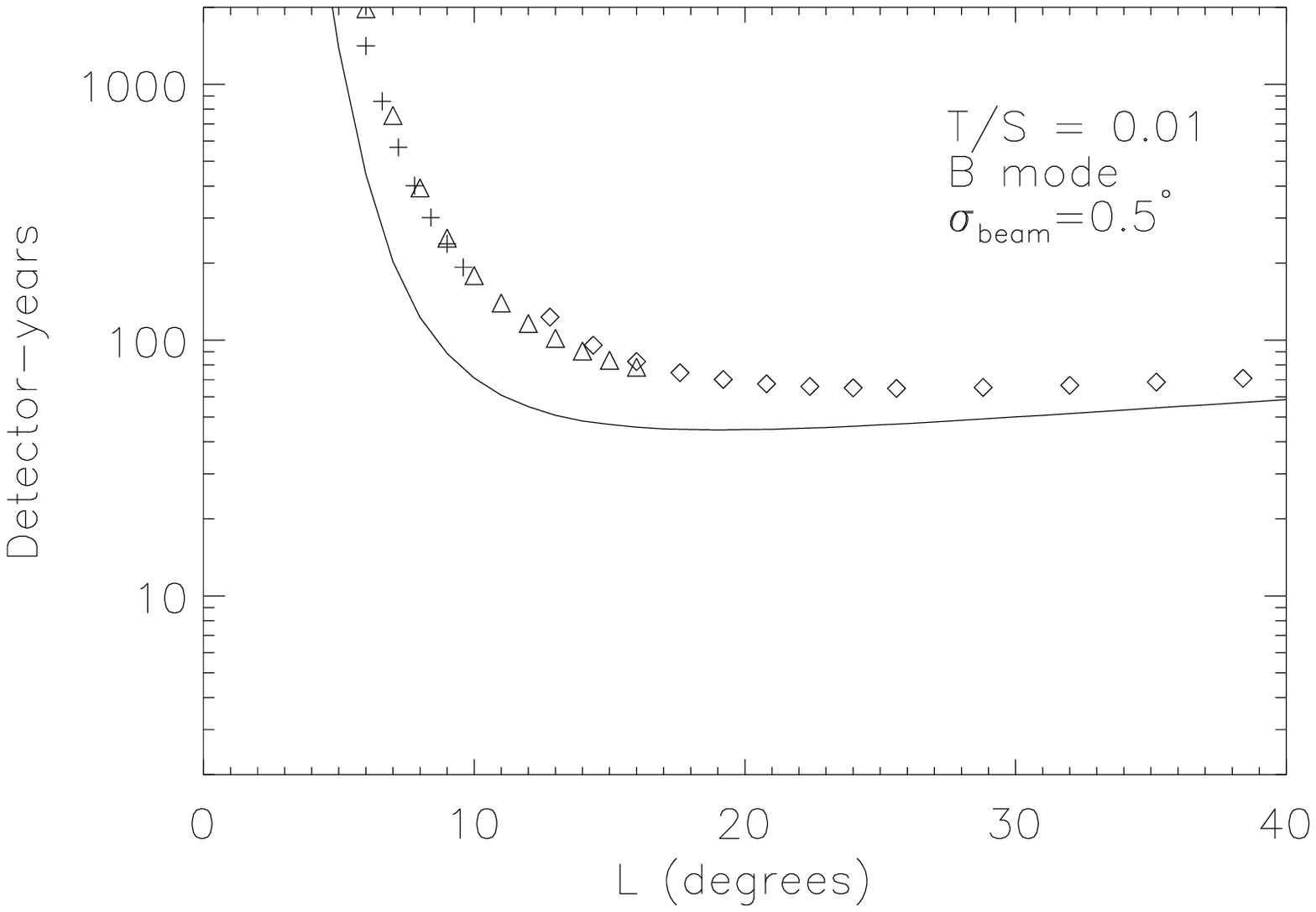}}
\caption{E and B component detectability for different beam sizes.
The results shown here all have $T/S=0.01$.  The symbols
have the same meanings as in
Figure \ref{fig:result1}, except for the pixel sizes: for a beam size
of 0.1$^\circ$, the different symbols represent
pixel sizes of $0.06^\circ,0.1^\circ,0.15^\circ$, while for
the 0.5$^\circ$ beam the symbols represent
$0.3^\circ,0.5^\circ,0.8^\circ$ pixels.}
\label{fig:beam}
\label{fig:resultlast}
\end{figure}

Once all of the above parameters have been specified,
we can compute
the E and B component power spectra with
CMBFAST, COBE-normalize \cite{cmbfast,bunnwhite}, and smooth with a Gaussian
beam $\sigma_{\rm beam}$.
We then use the formalism of Section \ref{sec:fisher} to calculate
the signal-to-noise ratios for the amplitude of the E and B components.
Since we are interested in determining the required
experimental parameters for a strong detection of the two
amplitudes, we fix the desired signal-to-noise at 3 and
solve numerically for the required weights.
The results are shown in Figures \ref{fig:result1} and \ref{fig:beam}.
For comparison, we also show the result
of the JKW approximation and, for
the case of the E component, the result of the ``enhanced'' JKW
approximation of equation (\ref{eq:jkwenh}).
Although most of our results were calculated with 
$\sigma_{\rm beam}=0.3^\circ$, we calculated the $T/S=0.01$
power spectra for $\sigma_{\rm beam}=0.1^\circ$ and $0.5^\circ$
as well to illustrate the effect of beam size on
our results.

In the Fisher matrix analysis, we analyze maps with several
different pixel sizes, as indicated
in the figure captions.  For the $0.3^\circ$ beam,
for example, the largest pixel size is 0.5$^\circ$,
which is somewhat larger than one would ideally like to use.
We use this large pixel size so that
we can investigate data sets with large sky coverage without
having to invert inconveniently large matrices.
As shown in the figures, there is overlap in map size
between different pixel sizes,
which gives a crude idea of the amount of information 
being lost when large pixel sizes are used.
In general,
the bulk of the information
is found on scales larger than the pixel scale; for
maps with a reasonable number of pixels,
the fact that we are undersampling the beam does
not appear to affect the results very much.

Note that the JKW approximation underestimates the difficulty 
of detecting the B component as expected.  For small
survey sizes, though, it overestimates the difficulty
of measuring the E component amplitude.  The enhanced JKW
approximation provides a better fit in this regime.

The detectability of the E component is quite flat as a function
of survey size even for fairly small
maps, but that of the B component is not.  
For the case
of a $0.3^\circ$ beam, the optimal survey size
for detecting
the B component is 22$^\circ$, as compared to 17$^\circ$ found via the JKW
approximation.  For $\sigma_\mathrm{beam}=0.5^\circ$, the optimum is
26$^\circ$, while the JKW approximation would give $19^\circ$.
For the $\sigma_\mathrm{beam}=0.1^\circ$ case, the
optimum survey size was too large to find with the Fisher-matrix
formalism.  The JKW optimum for this case occurs at $L=15^\circ$, with
a detection time of 30 detector-years; it is reasonable
to suppose that the true optimum is at even larger scales.

While it may not be necessary for an experiment to achieve the optimum
survey size, one would certainly like to stay well to the
right of the ``knee'' in Figures \ref{fig:result1} and \ref{fig:beam}.
For the case of a $0.3^\circ$ beam, the required detection time
is relatively flat for $L\gtrsim 12^\circ$ but rises sharply
at smaller $L$.  This suggests that a map of at least
$40\times 40$ $\sigma_{\rm beam}$-sized pixels should be made
in order to measure the B-component amplitude in such an experiment.
The experimental weight required for a 3-sigma B-component
detection in such an experiment would
be such that the B-component-signal--to--noise per pixel is approximately
unity.

\begin{figure}[t]
\centerline{\epsfxsize 4in\epsfbox{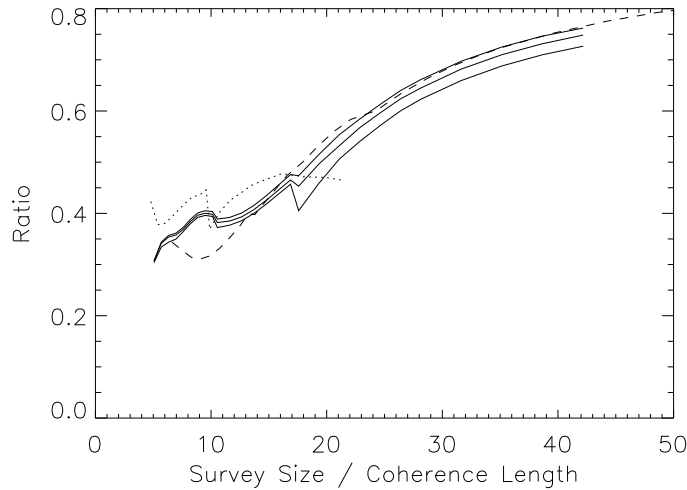}}
\caption{Comparison of JKW and Fisher matrix detectabilities.
The quantity on the vertical axis is the ratio of the required
B-mode detection time calculated using the JKW approximation to that calculated
from the Fisher matrix.  On the horizontal axis is the survey
size measured in units of the B-mode coherence length defined in the text.
The solid lines are the $\sigma_\mathrm{beam}=0.3^\circ$ models
from Figure \ref{fig:result1};
$T/S$ decreases from top to bottom.
The dotted and dashed lines are the $0.1^\circ$- and $0.5^\circ$-beam
models plotted in Figure \ref{fig:beam}.
The kinks in the curves are caused by the fact that different
pixel sizes were used in the maps.}
\label{fig:ratio}
\end{figure}

In order to determine whether
reionization made a significant
difference to our results, we examined a model in which
the optical depth to last scattering was $\tau=0.3$.
In order to keep the temperature power spectrum more or less the same,
we tilted the spectral index $n$ from 1 to 1.15.  As shown
in Figure \ref{fig:specs}, this makes a large difference
to the polarization power spectra, but most of the
difference is on larger scales than those to
which our degree-scale experiments are most sensitive.
The required detection times for this model were very similar
to those in the no-reionization case: for the E-component,
the difference was
less than 4\%, while for the B-component it ranged from 15\% down to 5\%
over the range of survey sizes shown in Figure \ref{fig:result1}.

\section{Discussion}
\label{sec:disc}

As we have seen in the previous section, the Fisher-matrix
formalism gives a slightly better detectability for the E-component
power spectrum amplitude than is predicted by the no-mixing
JKW approximation, at least in the case of small survey
sizes.  The reason for this is that some modes that
``should'' probe the B component are actually dominated by the
E component, increasing the number of independent
modes that can be used to estimate the E-component
amplitude.  
The enhanced JKW approximation (\ref{eq:jkwenh}) does surprisingly
well at characterizing the E-component detectability, confirming
that this ``extra mode'' explanation is indeed correct.

The main practical consequence of the enhanced
E-component detectability is that experiment
designers need not be terribly concerned about ensuring large
sky coverage: the detection requirements are fairly flat
as a function of survey size.
For the B component, on the other hand, survey size is quite
important.  The required observation time rises sharply as the
survey size decreases.  For a 0.3$^\circ$ beam, the ``knee''
in detectability occurs at a survey size of about 12$^\circ$.

As expected, the neglect of mixing in the JKW approximation causes
it to underestimate the difficulty of measuring the B component.
We might expect this underestimation to depend primarily on the
survey size, measured relative to some sort of coherence
length of the signal being looked for.
In Figure \ref{fig:ratio}, we  illustrate this
by plotting
the ratio of the detection time calculated using the JKW approximation
to the Fisher-matrix calculation.  On the horizontal axis is plotted
the survey size, measured in units of the B-component coherence length.
The coherence length is defined by $\theta_{\rm coh}=\pi/l_{\rm coh}$,
where the coherence multipole is 
\be
l_{\rm coh}\equiv{\sum_l (2l+1)l^2C_l^B\over\sum_l(2l+1)lC_l^B},
\ee
the weighted average multipole, with weights given by $l(2l+1)C_l^B$.
(Here $C_l^B$ is the beam-smoothed power spectrum as usual.)
All of the models discussed in the last section are plotted here.  
In general, the JKW approximation becomes good when the survey
size is many coherence lengths, but for $L\lesssim 20$ coherence
lengths, it underestimates the detection difficulty by a factor
of $\sim 2.5$.

It is important to note that the Fisher-matrix formalism used
in this paper gives the \textit{best} possible error bars
that can be achieved from a given experiment.  Even if, for example,
we expand the data set in some set of normal modes that minimize
E-B mixing \cite{ChiMa,TegCos,LewChaTur}, we cannot do any better
than the brute-force likelihood analysis of the entire data
set on which the Fisher-matrix formalism is based.
This is not to say, of course, that such methods are not useful.
They may reduce computation time, provide
insight into the nature of the E-B decomposition, and allow
filtered real-space maps of the E and B components to be made, for example.

Although the Fisher-matrix results represent the idealized
minimum possible error bars, they are likely to be quite close to
the errors achievable in real experiments.\footnote{Unless
foreground contamination, which is not treated in
the present work, proves to be a serious problem.}
After all, lossless or nearly lossless methods such 
as those that have been applied to recent temperature anisotropy
measurements \cite{boom,max} can be easily adapted to the polarization
case.  The other idealization that has been made in this analysis
is that only the amplitudes, and not the shapes, of the power
spectra are unknown.  This will of course not be true
in a real experiment (although, as long as CMB temperature
anisotropy experiments continue to return results consistent with
standard models, strong constraints on the shapes
will be available); however, for the first few
experiments to detect the E and B modes, a band-power estimate of the
power spectrum in a few bands will probably be sufficient and
may be expected to yield error bars similar in size to those
calculated by the Fisher-matrix formalism.

E-B
mixing is significant only on scales comparable to the size of the survey,
and therefore affects a relatively small number of 
modes \cite{TegCos,LewChaTur}.
It might therefore be surprising that mixing has as large an effect on
the B-component detectability as it does.  The main point to remember
in this regard is that the modes with wavelengths of order
the survey size 
are always the ones that are detected with the highest signal-to-noise.
(After all, the beam-smoothed $C_l^{B}$ is invariably a
sharply decreasing function of $l$, whereas the noise has a flat
power spectrum.)  The loss of these modes to mixing therefore
has a disproportionately large effect.

It should be noted that the criterion adopted for ``detectability''
of a component in this paper is that the amplitude of that
component's power spectrum can be measured with small fractional
uncertainty.  It is possible for a component to be ``detected''
(\textit{i.e.}, for the null hypothesis
that the component is absent to be ruled out) at high significance 
even if this criterion is not met.  For instance, if a single
mode is detected with high signal-to-noise, and if that
mode is known to be a pure B-mode, with no E-component contamination,
then the B-component will have been detected with high significance.
However, based on a single mode, the amplitude of the power spectrum
cannot be determined with fractional uncertainty less than $O(1)$,
so such an experiment would not meet the detectability criterion
considered in this paper.

This scenario could easily occur in an experiment designed
to measure the Stokes parameters in a thin ring \cite{Kea,Zal2,ChiMa}.
Because an accurate characterization of the power spectrum
amplitudes will be extremely important in interpreting
polarization results, and because a detection of the
polarization in many different modes is much more robust
than a detection in only a few, we have chosen to adopt
the stronger detectability criterion of this paper.

The formalism described in this paper has numerous applications.
Although we have considered only square maps, it can of course
be used to explore the effects of survey geometry on 
detectability of the two components.  It can also easily be adapted
to examine the degree to which inhomogeneous noise alters
detectability.  One might suppose, for instance, that
measuring the edges of the survey with high precision would help
in separating the two components, since the modes that are 
``ambiguous'' with respect to the E-B split tend to be supported
most strongly near the boundary \cite{LewChaTur}.
Finally, the formalism described herein may prove useful
in studies of weak gravitational lensing (\textit{e.g.}, \cite{lensing,lensing2}):
the shear induced by lensing is a spin-2 field, and the
mathematics is therefore quite closely analogous to the case
of CMB polarization.

\begin{acknowledgments}
The author wishes to thank Max Tegmark for useful conversations.  This
research was begun while the author was an ITP Scholar at the Institute
for Theoretical Physics at U.C.~Santa Barbara and has been supported
in part by NSF grant AST-0098048.
\end{acknowledgments}

\end{document}